\documentstyle[epsfig,11pt,twoside]{article}

\newcommand{\unit}[1]{\mbox{\ #1}}

\newcommand{\comment}[1]{}

\newcommand{\D}{\displaystyle}

\newcommand{\mt}{m_{\tau}}
\newcommand{\gnr}{\Gamma_{\tau \rightarrow \nu \pi}}
\newcommand{\ga}{\Gamma_{a_{1}}}

\newcommand{\be}{\begin{equation}}
\newcommand{\ee}{\end{equation}}

\begin{document}
\everymath={\displaystyle}
\thispagestyle{empty}
\vspace*{-2mm}
\thispagestyle{empty}
\noindent
\hfill TTP93--1A\\
\mbox{}
 \hfill  May 1994  \\
\vspace{0.3cm}
\begin{center}
  \begin{Large}
  \begin{bf}
ADDENDUM TO OUR PAPER:\\
``RADIATIVE TAU DECAYS\\
WITH ONE PSEUDOSCALAR MESON''\\
  \end{bf}
  \end{Large}
  \begin{large}
   Roger Decker\footnote{rd@ttpux1.physik.uni-karlsruhe.de} and
   Markus Finkemeier\footnote{mf@ttpux2.physik.uni-karlsruhe.de}
   \\[5mm]
    Institut f\"ur Theoretische Teilchenphysik\\
    Universit\"at Karlsruhe\\
    76128 Karlsruhe\\ Germany\\
  \end{large}
  \vspace{0.5cm}
  {\bf Abstract}
\end{center}
\begin{quotation}
\noindent
In a previous paper we have calculated the decay
$\tau\to\pi\nu_\tau\gamma$.
We used a given relative sign between
the internal bremsstrahlung and the structure dependent radiation,
which has been used by several other authors before.
However, we believe that there are good arguments why the opposite
sign is the physical one. We briefly discuss these arguments and then
investigate how the predictions for the decay
$\tau\to\pi\nu_\tau\gamma$ are affected by this sign.
\end{quotation}
%
%
In \cite{Dec93} we have calculated the radiative decay $\tau\to\pi(K)
\nu_\tau \gamma$.
In that paper we used structure dependent form factors $F_V$ and $F_A$
with
negative values at  $t=0$, ie.\ $F_V(t) < 0$, $F_A(0) < 0$ (see
Eqs.~(39) and (59) in \cite{Dec93}). In other words, for the
relative sign $s$
between
the structure dependent radiation SD and  the internal
bremsstrahlung IB which is defined by
\be
     s := \frac{F_V(0)  f_\pi}{| F_V(0)  f_\pi|}
\ee
we used $s=-1$.
We took this sign from \cite{Bae68}, where a detailed
derivation of the amplitude for $\pi\to e \nu_e \gamma$ is given.
This sign agrees
with the one in \cite{Bro64,Bry82}.
(In comparing with the
literature, some care is needed because of varying
conventions for the form factors and  for $\epsilon^{0123}$.)
The sign $s=-1$ in \cite{Bae68}, however, was chosen arbitraryly,
since no explicit model predicting the phases of $f_\pi$ and $F_V$ was
considered \cite{Pes94}.

Note that the value of $s$
affects the interference between SD and IB only.
In the case of the radiative pion decay $\pi\to e\nu_e\gamma$,
this interference is vanishingly small, so up to now a direct
measurement of $s$ was not possible.

However, the sign $s$ can be determined by using specific models or
symmetry relations between the phases of the form factors $F_V$ or
$F_A$ and that of other observables. For the vector form factor,
it is shown in \cite{Ter72} that it is related to the anomalous form
factor $H$ in $K_{l4}$ decays and that the available data for $H$
and $SU(3)$ symmetry relations
require $s=+1$.
In \cite{Gas84,Bij92}, the processes $\pi(K)\to l \nu_l \gamma$ are
calculated within the framework of chiral perturbation theory, also
resulting in $s=+1$.

Therefore we believe that most probably $s=+1$ is the physical choice.
We will discuss below how the predictions for
$\tau\to\pi\nu_\tau\gamma$ change under $s = -1 \to +1$.
In fact it will turn out that a measurement of this decay yields
$s$.

In \cite{Dec93} we compared our results to those of
\cite{Kim80,Iva90}. We stated that we agreed with \cite{Kim80}
if we used their
particle data and their parametrization of the form factors, but that we
did not reproduce the numerical results of \cite{Iva90}.
Now it appears that this is due to the fact that the authors in
\cite{Kim80} use $s=-1$ while those in \cite{Iva90} have $s=+1$.
With the latter choice and the parametrization of \cite{Iva90} we fully
reproduce their results.

\begin{figure}
\caption{Pion-photon invariant mass spectrum of the
   decay $\tau\to\nu\pi\gamma$ (using the standard parameter set)}
\label{figb6}
\begin{center}
\fbox{   \epsfig{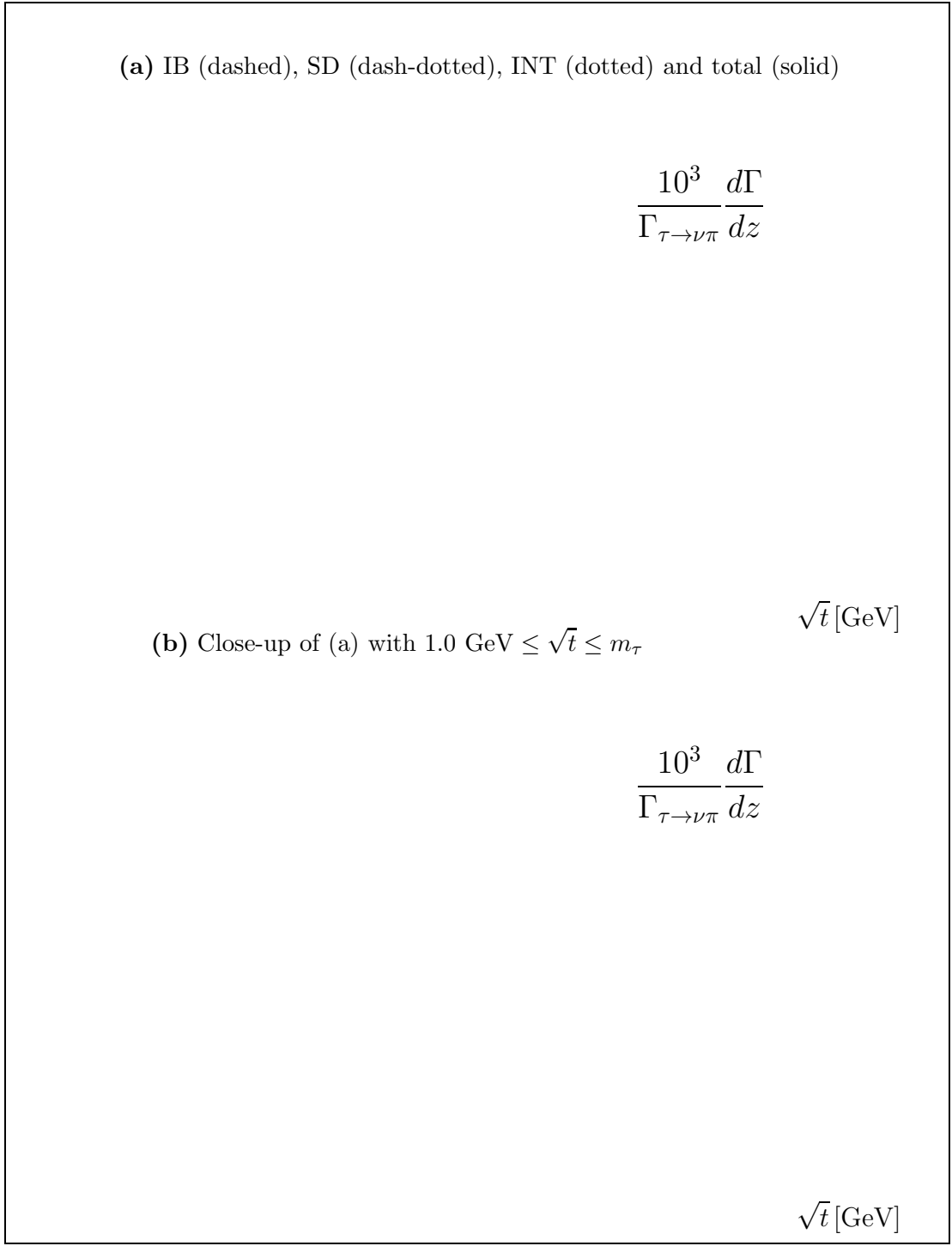}}
\unitlength1.0cm
\begin{picture}(0,0)
   \put(-13,18.1){\makebox(0,0)[l]{{\bf (a)} IB (dashed), %
     SD (dash-dotted), INT (dotted) and total (solid)}}
   \put(-12.5,9.2){\makebox(0,0)[l]{{\bf (b)} Close-up of (a) %
   with $\D 1.0 \unit{GeV} \leq \sqrt{t} \leq \mt$}}
  \put(-2.5,9.6){\makebox(0,0)[l]{\large  $\D \sqrt{t}\, \mbox{[GeV]} $ }}
  \put(-5,16.){\makebox(0,0)[l]{\Large $ \D\frac{ 10^{3}}{ \gnr}%
  \frac{d\Gamma}{dz}$}}
  \put(-2.5,0.3){\makebox(0,0)[l]{\large  $\D \sqrt{t}\, \mbox{[GeV]} $ }}
  \put(-5,7.){\makebox(0,0)[l]{\Large $ \D\frac{ 10^{3}}{ \gnr}%
  \frac{d\Gamma}{dz}$}}
%
\end{picture}
  \end{center}
\end{figure}
\begin{figure}
\caption{Pion-photon invariant mass spectrum in the $a_1$ mass region,
using $\ga = 250$~(dashed), $400$ (solid) and $650 \unit{MeV}$
(dotted)}
\label{figb7}
\begin{center}
\fbox{\epsfig{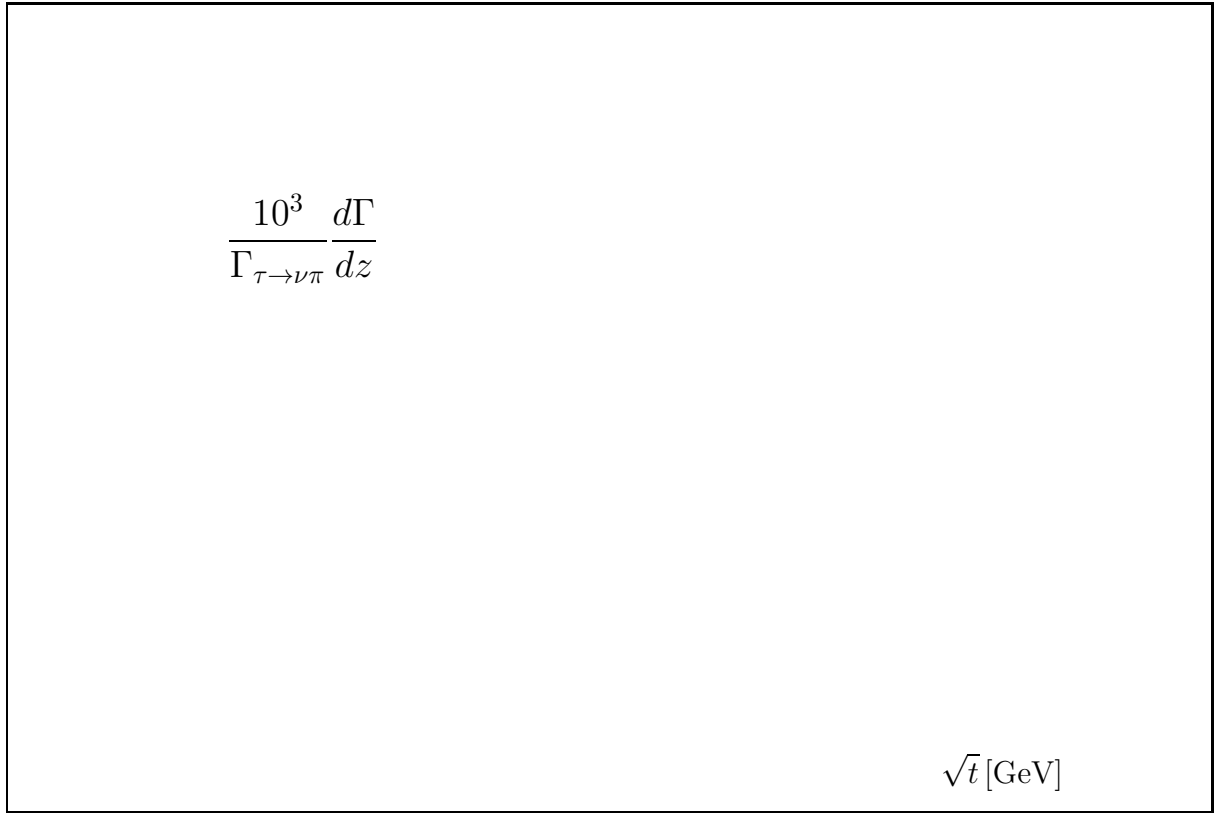}}
\unitlength1.0cm
\begin{picture}(0,0)
%
%
  \put(5.,0.9){\makebox(0,0)[l]{\large  $\D \sqrt{t}\, \mbox{[GeV]} $ }}
  \put(-4.5,8.){\makebox(0,0)[l]{\Large $ \D\frac{ 10^{3}}{ \gnr}%
  \frac{d\Gamma}{dz}$}}
%
%
%
\end{picture}
  \end{center}
\end{figure}

Now we will discuss the changes which occur in \cite{Dec93} if we take
$s=+1$.
We have to replace $F_V(0)$ and $F_A(0)$ of \cite{Dec93} by
\begin{eqnarray}
   F_V(0)  = + 0.0270 \qquad F_A(0) = +0.0116
\end{eqnarray}
Then the rates for the internal bremsstrahlung $\Gamma_{IB}$ and for
structure dependent radiation
$\Gamma_{SD}$ remain unaffected, while the interference term
$\Gamma_{INT}$ changes its sign. So in the numerical discussion in
Sec.~5 of \cite{Dec93}, the sign of all interference contributions
(IB-V, IB-A and
INT, cf.\
Eqns.~(27)) have to be reversed.
This concerns tables 1, 2, 4.
The interference becomes
destructive, such that the difference between the pure internal
bremsstrahlung IB and the full answer $\Gamma_{total}$ is always
reduced.
Similarly most of the numbers in table 3 are somewhat reduced, from
about $2.8$ down to values of about $1.9$.
Also in the photon spectrum (Fig.~4(a,b)), the interference INT (dotted)
must be subtracted rather than added to IB and SD, if $s=+1$. This
brings the total spectrum even closer to the pure internal
bremsstrahlung (IB), especially for large $x$.
While Fig.~4(c) is not
affected at all, in Fig.~4(d) the numbers on the vertical axis change
their sign.
Fig.~5 does not change much, again the total rate comes closer to the
internal bremsstrahlung.

Let us now discuss the most important changes, which occur in the
pion-photon invariant mass spectrum.  Using $s=+1$
we obtain  Figs.~1 and 2, which are to
be compared with Figs.~6 and 7 of \cite{Dec93}.
The destructive
interference substantially diminishes the total rate  above the
$\rho$ resonance peak at $t=m_\rho^2$ (Fig.~1(a)). The destructive
interference
contribution decreases strongly if $t$ approaches $m_{a_1}^2$ from
below, such that with $s=+1$ a $a_1$ resonance peak becomes visible
(Fig.~1(b)).
As is shown in Fig.~2, the width $\Gamma_{a_1}$ can be measured quite
well in this spectrum if $s=+1$.

In order to remain brief, we will not give
all the numbers for the decay rates and plots of the spectra for
$s=+1$ in this addendum. They can, however, be found in
\cite{Fin94}.

To conclude let us repeat our main results.
First we find that the decay $\tau\to \pi \nu_\tau \gamma$
allows for direct measurement of the relative sign $s$ between the
internal bremsstrahlung and the structure dependent radiation. Second,
if $s=+1$, which is most probably the physical choice, then the
difference between the pure internal bremsstrahlung and the physical
spectrum is very small in most of the phase space, because of the
destructive interference between the internal bremsstrahlung and the
structure dependent radiation. Third, if $s=+1$, a measurement of the
pion-photon invariant mass spectrum allows for a measurement of
$\Gamma_{a_1}$.




\begin{thebibliography}{[99]}
%
%
\bibitem{Dec93}
R. Decker and M. Finkemeier, Phys. Rev. {\bf D 48} (1993) 4203
%
\bibitem{Bae68}
P. de Baenst and J. Pestieau, Nuovo Cimento {\bf 53 A} (1968) 407
%
\bibitem{Bro64}
S. G. Brown and S. A. Bludman, Phys. Rev. {\bf 136} (1964) B1160
%
\bibitem{Bry82}
D. A. Bryman, P. Depommier, C. Leroy, Phys. Rep. {\bf 88} (1982) 151
%
\bibitem{Pes94}
J. Pestieau, personal communication, 1994
%
\bibitem{Ter72}
M.V. Terent'ev, Yad. Fiz. {\bf 16} (1972) 1044, Sov. J. Nucl. Phys.
{\bf 16} (1973) 574
%
%
\bibitem{Gas84}
J. Gasser and L. Leutwyler, Ann. Phys. (N.Y.) {\bf 158} (1984) 142
%
\bibitem{Bij92}
J. Bijnens, G. Ecker and J. Gasser, Nucl.Phys. {\bf B 396} (1993) 81
%
\bibitem{Kim80}
J. Kim and L. Resnick, Phys. Rev. {\bf D 21} (1980) 1330
%
\bibitem{Iva90}
Yu.P. Ivanov, A.A. Osipov, M.K. Volkov, Phys.~Lett.~{\bf B 242}
(1990) 498; Yu.P. Ivanov, G.V. Mitselmakher, A.A. Osipov,
Z. Phys. {\bf C 50} (1991) 113
%
\bibitem{Fin94}
M.~Finkemeier, ``Radiative Corrections to the Decay
$\tau\to\pi\nu_\tau$'', Ph.~D. thesis, Universit\"at Karlsruhe, 1994,
to be published by Verlag Shaker, Aachen, Germany.
(Please contact the author if you have any problems in obtaining a
copy.)
%
\end{thebibliography}
\end{document}